\documentclass[
  reprint,
  amsmath,amssymb,
  aps,
  prab,
  floatfix
]{revtex4-2}

\usepackage{graphicx}
\usepackage{dcolumn}
\usepackage{bm}
\usepackage{etoolbox}
\usepackage{makecell}
\usepackage{multirow}
\emergencystretch=\maxdimen
\usepackage{amsmath}
\newcommand{\xl}{\textcolor{black}}
\newcommand{\xlu}{\textcolor{black}}

\usepackage[caption=false]{subfig}
\usepackage{hyperref}    
\hypersetup{
    colorlinks=true,
    linkcolor=blue,
    citecolor=blue,
    urlcolor=blue
}
\usepackage{siunitx}      
\begin{document}

\title{Analytical and Numerical Studies of Dark Current in Radiofrequency Structures\\for Short-Pulse High-Gradient Acceleration}

\author{Gaurab Rijal}
\affiliation{Northern Illinois University, DeKalb, Illinois  60115, USA}

\author{Michael Shapiro}
\affiliation{Northern Illinois University, DeKalb, Illinois  60115, USA}

\author{Xueying Lu}
\email{xylu@niu.edu}
\altaffiliation{also at Argonne National Laboratory, Lemont, Illinois 60439, USA}
\affiliation{Northern Illinois University, DeKalb, Illinois  60115, USA}

% \maketitle

% \date{\today}

 \begin{abstract}
High-gradient acceleration is a key research area that could enable compact linear accelerators for future colliders, light sources, and other applications. In the pursuit of high-gradient operation, RF breakdown limits the attainable accelerating gradient in normal-conducting RF structures. Recent experiments at the Argonne Wakefield Accelerator suggest a promising approach: using short RF pulses with durations of a few nanoseconds. Experimental studies show that these $\mathcal{O}\!\bigl(1\,\text{ns}\bigr)$ RF pulses can mitigate breakdown limitations, resulting in higher gradients. For example, an electric field of nearly 400~MV/m was achieved in an $X$-band photoemission gun driven by 6-ns-long RF pulses, with rapid RF conditioning and low dark current observed. Despite these promising results, the short-pulse regime remains an under-explored parameter space, and RF breakdown physics under nanosecond-long pulses requires further investigation. In this paper, we present analytical and numerical simulations of dark current dynamics in accelerating cavities operating in the short-pulse regime. We study breakdown-associated processes spanning different time scales, including field emission, multipacting, and plasma formation, using simulations of the $X$-band photogun cavities. The results reveal the advantages of using short RF pulses to reduce dark current and mitigate RF breakdown, offering a path toward a new class of compact accelerators with enhanced performance and reduced susceptibility to breakdown.
\end{abstract}

\maketitle

\section{Background and Introduction}
\label{sec:introduction}
Achieving a high accelerating gradient is crucial for compact linear accelerators, but the presence of radiofrequency (RF) breakdown poses a significant challenge. Recently, acceleration with RF pulses on the order of a few nanoseconds has emerged as a promising approach for overcoming long-standing limitations in the gradient achievable by mitigating the impact of RF breakdown~\cite{dobert_tech,Dobert2007HighPT,dyu,wuensch_PAC,tan-2022,shao-ipac2022,mtm_prab,freemire-2023}. These short RF pulses have been generated by various approaches including RF pulse compressors~\cite{ivanov,samsonov,gu_prab2025}, and power extractors~\cite{syratchev,CAPPELLETTI,picard-2022,peng_ipac2019,shao-2020} based on the structure wakefield acceleration (SWFA) concept~\cite{Lu:2022oin}. SWFA is one promising advanced accelerator concept, where a high-charge drive beam excites wakefield in metallic or dielectric structures in vacuum, and then the excited wakefield can be used to accelerate a main beam at high gradients. There are two schemes of SWFA: collinear wakefield acceleration, where the drive and main beams travel in the same structure, and two-beam acceleration (TBA), where the two beamlines are decoupled. In the TBA scheme, a power extractor is designed to extract the wakefield generated by the drive beam, often a train of high-charge electron bunches. Each electron bunch in the train excites a short RF pulse upon traversing the power extractor, and when the wakefield pulses generated from multiple electron bunches are coherently added, a combined RF pulse with high peak power can be produced. Using this approach, RF pulses with a duration of a few nanoseconds and a peak power of over 500~MW have been generated at $X$-band~\cite{picard-2022,peng_ipac2019}. These short RF pulses can then be transferred and coupled into accelerating structures, where gradients exceeding the limit in conventional accelerators (100~MV/m for $X$-band for example) have been demonstrated in a series of experiments~\cite{jing2022}.

\begin{figure*}[htbp]
    \centering
    \includegraphics[width=\textwidth]{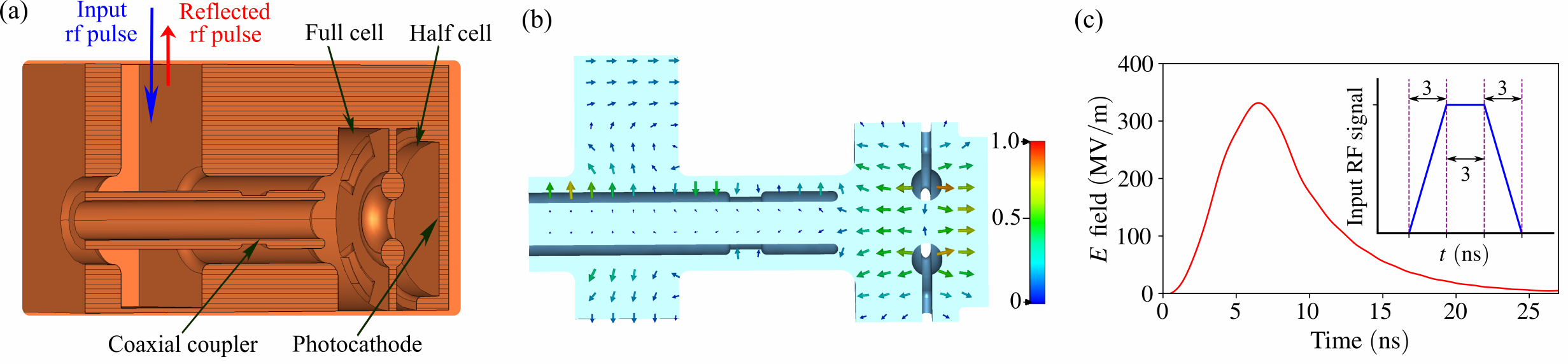}
    \caption{Model and RF design of the $X$-band photogun cavities driven by short RF pulses.
    (a) Cross-sectional view of the copper cavities along the central plane.
    (b) Normalized electric field \xl{magnitude} distribution at 11.7~GHz. 
    (c) Electric field on the photocathode surface for an input RF pulse with 200~MW peak power; the inset shows the time profile of the input RF signal, with a 3~ns rise, 3~ns flat top and 3~ns fall.}
    \label{fig:Efield_combined}
\end{figure*}

One of these experiments at the Argonne Wakefield Accelerator (AWA) facility successfully demonstrated an electric field approaching 400~MV/m on the photocathode surface in an $X$-band photogun with (1+1/2)-cell resonant cavities~\cite{tan-2022}. These cavities are powered by approximately 9~ns long RF pulses extracted from an eight-electron-bunch train with a total charge of about 400~nC using an $X$-band power extractor~\cite{peng_ipac2019}. Figure~\ref{fig:Efield_combined}(a) shows the photogun cavities, comprising a full cell and a half cell, and an input waveguide port. A coaxial coupler was designed to maintain good field symmetry at the photocathode. A directional coupler (not shown) was attached to the waveguide port to measure the input and reflected RF signals. The cavities operate in the TM$_{010, \pi}$ mode with a center frequency of 11.7~GHz, and the electric field distribution is shown in Fig.~\ref{fig:Efield_combined}(b). The input RF signal exhibits a time profile of approximately 3~ns rise, 3~ns flat top and 3~ns fall, as shown in Fig.~\ref{fig:Efield_combined}(c). This time profile is determined by the design of the power extractor and by the spacing and total length of the drive bunch train. The electric field at the center of the photocathode for an input pulse with a peak power of 200~MW is also shown in Fig.~\ref{fig:Efield_combined}(c).

\begin{figure}[t!]
    \centering
    \includegraphics[width=0.9\columnwidth]{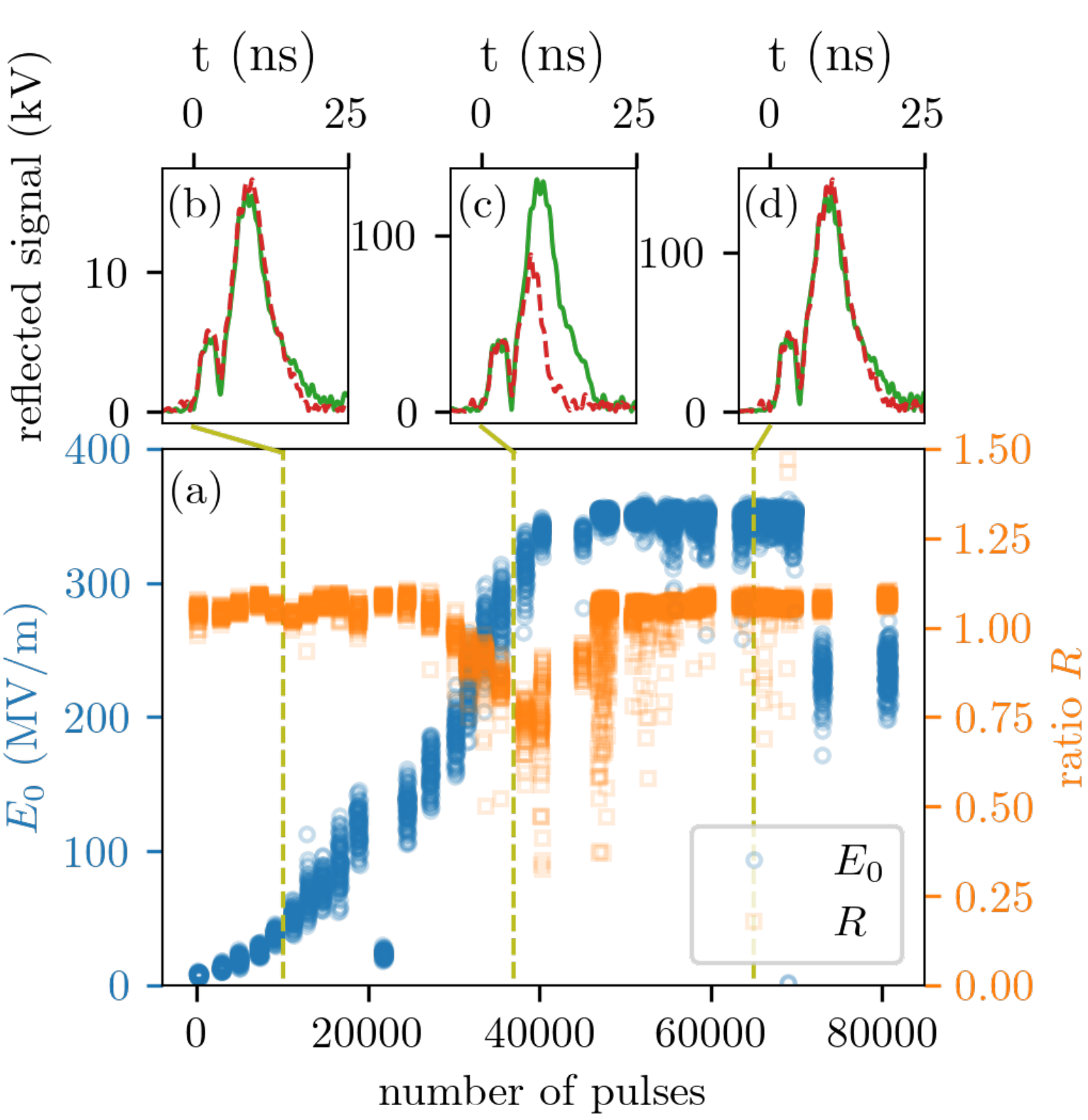}
    \caption{RF conditioning of the $X$-band photogun cavities. (a) Conditioning history of the peak electric field $E_0$ at the photocathode surface (blue), and the ratio of the measured to simulated peak reflected RF signal $R$ (orange). (b-d) \xl{Comparison of measured (dashed red) reflected RF signals with simulated signals (solid green, assuming no RF breakdown)} at the corresponding stages denoted by the vertical lines in Panel (a). This figure is reprinted with permission from Tan et al., Phys. Rev. Accel. Beams 25, 083402 (2022) under the terms of the Creative Commons Attribution 4.0 International license.}
    \label{fig:conditioning}
\end{figure}

The encouraging high gradient demonstrated experimentally points to the need for a deeper physics understanding of RF breakdown processes in the under-explored parameter space of $\mathcal{O}\!\bigl(1\,\text{ns}\bigr)$ pulses. An empirical scaling law~\cite{grudiev-2009}, established from breakdown studies in klystron-powered accelerating structures with $\mathcal{O}\!\bigl(100\,\text{ns}\bigr)$ RF pulse durations, relates the breakdown rate (BDR) to the accelerating gradient $E_a$ and the pulse length $t_p$ as $E_a^{30} \times t_p^5 / \text{BDR} = \text{constant}$, suggesting that higher gradients can be achieved with shorter pulses. However, recent experiments using $\mathcal{O}\!\bigl(1\,\text{ns}\bigr)$ long pulses revealed different scaling~\cite{mtm_prab,shao-ipac2022} and distinct dark current signatures~\cite{tan-2022}. During the RF conditioning of the short-pulse $X$-band photogun, very low dark current was observed. Despite monitoring with both a Faraday cup and an integrating current transformer, no clear signal was detected, suggesting an upper limit of approximately 1~pC on the dark-current charge \xlu{at the location of the monitors~\cite{tan-2022}}. Dark-current-induced beam loading was also observed in this experiment, as shown in Fig.~\ref{fig:conditioning}(a). The ratio $R$ of the measured and simulated peak reflected RF signals was close to unity in the absence of RF breakdown or dark current [Figs.~\ref{fig:conditioning}(b) and~\ref{fig:conditioning}(d)], but dropped below unity during a stage where beam loading occurred [Fig.~\ref{fig:conditioning}(c)]. These observations call for further investigation of the dark current dynamics in the short-pulse regime. 

Dark current refers to the unwanted flow of electrons in the absence of an injected beam~\cite{Chao:1490001}, and it can coexist with the primary beam and cause instabilities, power losses, and structural damage due to electron bombardment~\cite{Wilson2001}. The growth of dark current may also contribute to RF breakdown, which typically manifests as vacuum arcs that can disrupt RF power flow and may even cause irreversible damage to accelerating structures. RF breakdown is strongly influenced by the localized electromagnetic field on the inner surfaces of the structure, as well as by interactions between these fields and dark electrons~\cite{simakov-2018}. A range of physical processes across various timescales are associated with dark current dynamics.

A common source of dark electrons is field emission from surface regions with high local fields~\cite{Jensen2008, M_Jimenez_1994}. The local field $E_{\text{local}}$ is often significantly higher than the applied macroscopic field $E_{\text{applied}}$ due to surface irregularities, and can be expressed as $E_{\text{local}} = \beta E_{\text{applied}}$, where $\beta$ is the field enhancement factor. Material properties such as work function, roughness, and composition can strongly influence field emission~\cite{Fowler_Nordheim1928, Alpert1964, NOREM2005510, Ahearn1936}.  Field emission commonly occurs in high-field regions, where pulsed heating -- localized surface temperature rise via Joule heating -- can vaporize emitting sites, triggering explosive electron emission and surface modification, which may in turn further enhance field emission~\cite{wang-1997-a}. Electron multipacting is another important mechanism contributing to dark current growth~\cite{Vaughan1988,BIENVENU19921,Larciprete2013,Valizadeh2014,Han2008SinglesideEM,Mori2021DLC,Power2004,Yater2023,Haoran_2019}. In this process, emitted electrons from the surface interact with the RF field and may return to the surface again, releasing secondary electrons. Under certain resonant conditions, this process becomes self-sustaining and can lead to exponential electron multiplication. As the electron density increases, a plasma may form~\cite{Neuber1999,Valfells2000,Sun2018}.The resulting plasma can short-circuit RF structures, trigger further ionization, and escalate into arcing.

We report analytical and numerical studies on the dynamics of dark current in the short-pulse acceleration regime, using the $X$-band cavities shown in Fig.~\ref{fig:Efield_combined} as a representative case. Section~\ref{sec:analytical} presents an analytical theory of electron multipactor, where we identify the multipacting resonance conditions from electron trajectory calculations and evaluate the secondary electron yield (SEY). Section~\ref{sec:sims} describes simulations of field emission and multipacting processes using the \textsc{cst}~\cite{CST} Particle Studio, analyzing the impact of pulse shape, pulse length, and structure gradient on dark current growth, with a focus on comparing the long- and short-pulse regimes. Section~\ref{sec:impact} discusses the formation of multipacting electron clouds and their influence on RF pulse distortion. Finally, Sec.~\ref{sec:conclusions} summarizes the key findings and outlines future directions for high-gradient short-pulse RF acceleration research.

\section{Theory of Electron Multipacting}\label{sec:analytical}
In this section, we present an analytical treatment of electron multipacting by introducing a model which incorporates spatial variations in the electric and magnetic fields of a modified pillbox mode to capture electron motion near the cavity sidewalls. With this analytical model, we can precisely identify multipacting resonance modes in cylindrical accelerating cavities.

Previous analytical models have explored electron multipacting in various RF configurations~\cite{Kishek1998,Sazontov2011,pzhang,Shemelin2013, Lyneis1977,GonzalezIglesias2023NonresonantUM}. For example, Ref.~\cite{Kishek1998} presents a theory of single-sided multipacting on a dielectric surface with an RF electric field parallel to the dielectric surface and a DC electric field perpendicular to the surface. Ref.~\cite{Sazontov2011} investigates the effect of the RF magnetic field on multipacting on dielectric surfaces. These results
are important for evaluating multipacting behavior in dielectric windows. Multipacting on metallic surfaces of accelerator structures has also been studied. Ref.~\cite{pzhang} investigates two-sided multipacting between two metal electrodes. Ref.~\cite{Shemelin2013} analyzes the single-sided multipacting on cavity sidewalls under the influence of crossed fields. The study identifies electron trajectories that return to the emission surface and contribute to sustained discharge.

In this work, we investigate electron multipacting in crossed RF fields near the sidewall of a pillbox cavity, focusing on closed electron trajectories that originate from and return to the sidewall. These trajectories are of particular interest, as they are resonant with the RF field and are expected to play a dominant role in the multipacting process. Our analysis includes an axial electric field \(E_z\), a radial electric field \(E_r\), and an azimuthal magnetic field \(B_\theta\). The radial component \(E_r\) arises near the cavity sidewall due to the presence of the cavity iris. We modify the field distribution in the pillbox cavity to approximate conditions near the sidewall region of the \(X\)-band photogun cavities. We assume an axisymmetric field distribution, so electron motion is confined to the \(r\)–\(z\) plane. This assumption holds as long as any asymmetries or perturbations that could induce out-of-plane motion remain negligible. These approximations yield predictions in good agreement with simulations of the realistic cavity, as presented in Sec.~\ref{sec:pic_mp}. We begin by calculating secondary electron trajectories and then evaluate the SEY under both single-trip and multipacting resonance conditions.

\begin{figure}[t]
    \centering
    \includegraphics[width=0.75\columnwidth]{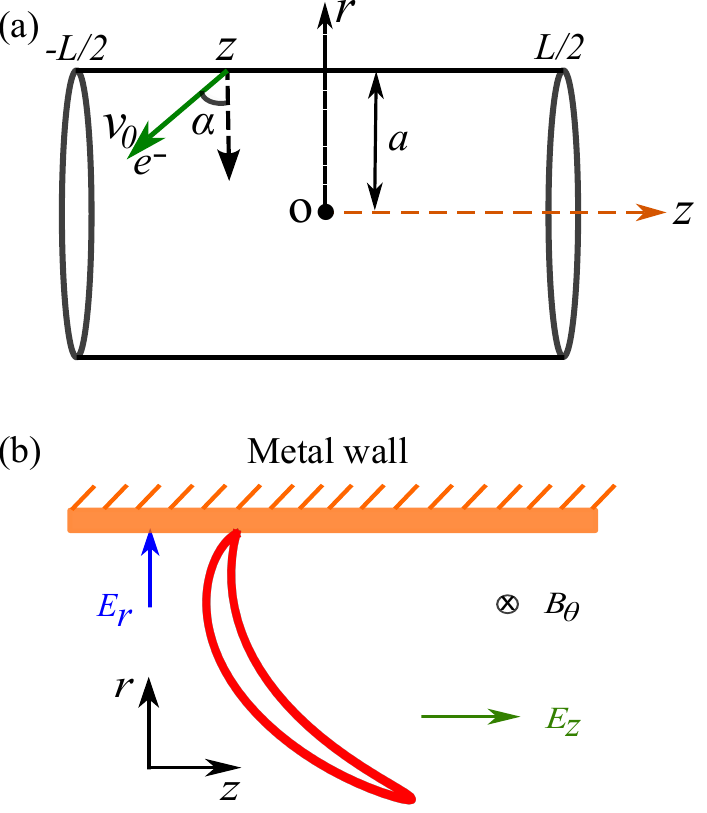}
    \caption{Schematic representation of the multipactor analysis setup. (a) Simulation geometry used for electron trajectory evaluation. (b) Crossed RF field configuration showing an electron trajectory under combined $E_r$, $E_z$, and $B_\theta$ fields. The parameters $v_0$ (initial velocity) and $\alpha$ (emission angle) are defined at emission. $r$ and $z$ denote the radial and longitudinal coordinates, respectively, while $a$ and $L$ are the cavity radius and length, respectively.}
    \label{fig:combined_crossfield}
\end{figure}

\subsection{Secondary electron trajectory calculations}
Under the assumptions outlined above, we investigate electron multipacting near the sidewall of a pillbox cavity. The cavity has a radius of \(a = 9.49\)~mm and a length of \(L = 6.50\)~mm, corresponding to the full-cell dimensions of the \(X\)-band photocathode cavities. The configuration used for the electron trajectory calculations is illustrated in Fig.~\ref{fig:combined_crossfield}. The origin, \(r = z = 0\), is defined at the center of the pillbox cavity. An electron is emitted from the sidewall at \(t = 0\) with an emission angle \(\alpha_0\), initial velocity \(v_0\), and initial RF phase \(\phi_0\). The initial axial position is \(z(t = 0) = z_0\), and the radial distance from the sidewall is \(\Delta r(t = 0) = r(t = 0) - a = 0\). We assume a steady-state \xl{RF} field, with the axial electric field amplitude at the cavity center, \(E_{z0}\), \xl{treated as constant}. The equations of motion for the electron are:
\begin{align}
    m\frac{dv_z}{dt} &= -e E_{z} - e v_r B_\theta,
    \label{eq:axial_motion} \\
    m\frac{dv_r}{dt} &= -e E_r + e v_z B_\theta,
    \label{eq:radial_motion}
\end{align}
where $v_r = dr/dt$ is the radial velocity, $v_z =dz/dt$ is the axial velocity, and $e$ and $m$ are the charge and mass of the electron, respectively.

\begin{figure}[t]
    \centering
    \includegraphics[width=0.9\columnwidth]{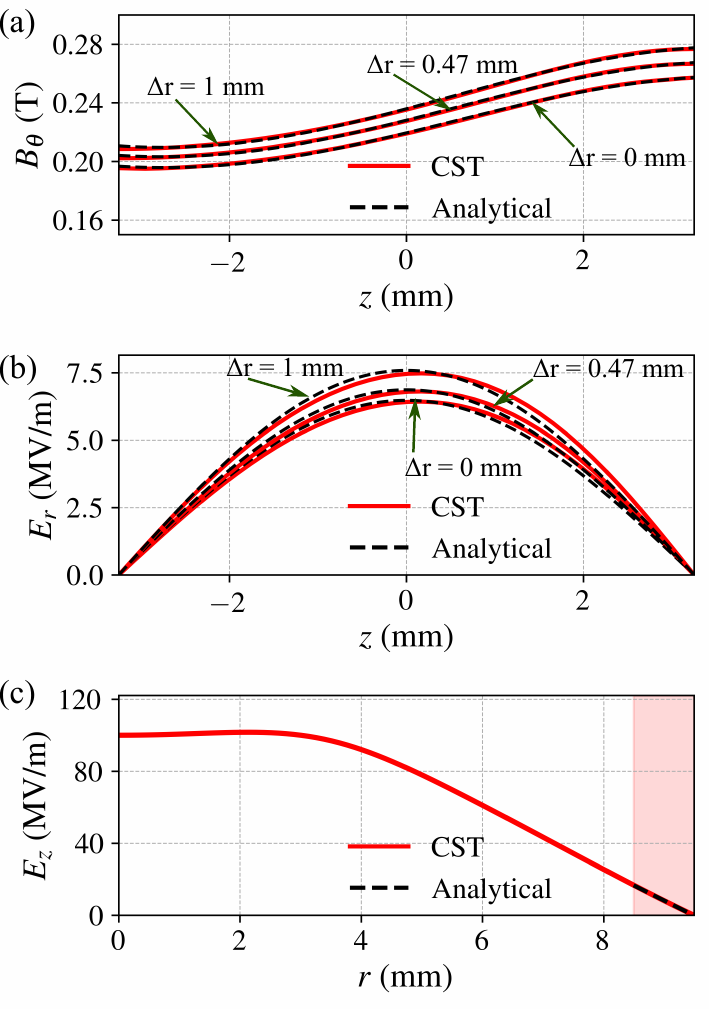}
 \caption{Comparison of RF field amplitudes from \textsc{cst} simulations and analytical approximations for the full-cell cavity with radius \( a = 9.49 \)~mm and length \( L = 6.50 \)~mm. The field amplitudes, \( B_\theta \) in (a), \( E_r\) in (b), and \( E_z\) in (c), are all scaled to an axial field amplitude of \( E_{z0} = 100 \)~MV/m at the cavity center. Solid red curves represent \textsc{cst} simulation data, and black dashed lines indicate the analytical field amplitudes used in the model. The shaded region in (c) indicates the radial range near the sidewall where the analytical field approximation is applied.}
    \label{fig:combined_field}
\end{figure} 

As secondary electrons remain localized near the cavity sidewall, we approximate the local RF fields as linear in $\Delta r = r - a$, the radial displacement from the sidewall.
 The radial electric field \(E_r\) is given by:
 \begin{equation}
E_r(\Delta r, z) = E_{r_0} \left(1 + S \Delta r \right) \cos\left(\frac{\pi z}{L}\right) \cos(\omega t + \phi_0),
\label{eq:Er_field}
\end{equation}
where $E_{r0}$ is the peak amplitude of $E_r$, \( \omega \) is the angular frequency of the RF field, and \xl{\( S = (dE_r/dr)/E_{z0}\)}, fitted from the simulated field distribution.
 The axial electric field \(E_z\) is:
\begin{equation}
E_z(\Delta r, z) = E_{z0} P \Delta r \cos(\omega t + \phi_0),
\label{eq:Ez_field}
\end{equation}
 where $E_{z0}$ again denotes the $E_z$ field amplitude at the cavity center, and \( P = (\mathrm{d}E_z/\mathrm{d}r)/E_{z0} \)\,. The azimuthal magnetic field \(B_\theta\) is modeled as:
\begin{multline}
B_\theta(\Delta r, z) = 
\left[ (B_0 + S_1 \Delta r) + (B_1 + S_2 \Delta r) \right. \\
\left. \times \sin\left(\frac{\pi z}{L} + \zeta \right) \right] \sin(\omega t + \phi_0),
\label{eq:Btheta_field}
\end{multline}
where the coefficients \( B_0 \), \( B_1 \), and \( \zeta \) are obtained by fitting to the simulated \( B_\theta \) distribution. \( S_1 = dB_0/dr \) and \( S_2 = dB_1/dr \) represent the radial gradients of the respective components near the sidewall. A comparison between the \textsc{cst}-simulated field in the full cell of the $X$-band cavities and the analytical approximations in Eqs.~\ref{eq:Er_field}--~\ref{eq:Btheta_field} is shown in Fig.~\ref{fig:combined_field}.

\xl{The above model with approximate field distributions uses a small set of fitted parameters and can be rapidly adapted to structures with similar field profiles.} While we employ the above field distributions to calculate secondary electron trajectories for the remainder of this section, we have also developed a simplified theory neglecting the radial variation of $E_z$ and $B_\theta$. This approximation yields closed-form expressions for multipacting resonance conditions, useful for rapid design estimates. The simplified theory, presented in Appendix~\ref{sec:simplified}, agrees well with the full model at low fields. For accurate trajectory tracking and SEY evaluation, however, we retain full radial dependence in all RF field components.

\subsection{\xl{Single-trip} SEY calculations}
\xl{We define a single-trip multipacting trajectory as the initial excursion from emission to the first return impact on any surface, regardless of RF period count elapsed or whether the electron returns to its emission site. This contrasts with the multi-cycle resonances in Sec.~\ref{sec:multi-resonance}, where repeated resonant electron impacts occur over multiple RF cycles.} Here we evaluate the SEY, $\delta$, as the number of secondary electrons emitted per incident electron, after a single emission–return trip.

The SEY depends on both the impact energy \( W_i \) and the angle of incidence \( \theta_i \). We use Vaughan’s empirical model~\cite{Vaughan1993} to evaluate $\delta$:
\begin{equation}
\delta = \delta_m \left[ w \exp(1 - w) \right]^k, \quad \text{where} \quad w = \frac{W_i}{W_m}.
\label{eq:vaughan}
\end{equation}
Here, \( \delta_m \) is the maximum yield, and \( W_m \) is the corresponding impact energy at which this maximum occurs. Both parameters increase with the angle of incidence $\theta_i$ according to:
\begin{equation}
\delta_m = \delta_{\text{max}} \left( 1 + \frac{\theta_i^2}{2\pi} \right), \quad
W_m = W_{\text{max}} \left( 1 + \frac{\theta_i^2}{2\pi} \right),
\label{eq:delta_params}
\end{equation}
where \( \delta_{\text{max}} = 2.1 \) and \( W_{\text{max}} = 150~\text{eV} \) are the values for normal incidence. The exponent \( k \) in Eq.~\ref{eq:vaughan} depends on the normalized energy \( w \) and is defined piecewise:
\begin{equation}
k =
\begin{cases}
0.56 & \text{if } w < 1, \\
0.25 & \text{if } w > 1.
\end{cases}
\label{eq:k_cases}
\end{equation}

\begin{figure}[t]
    \centering
    \includegraphics[width=\columnwidth]{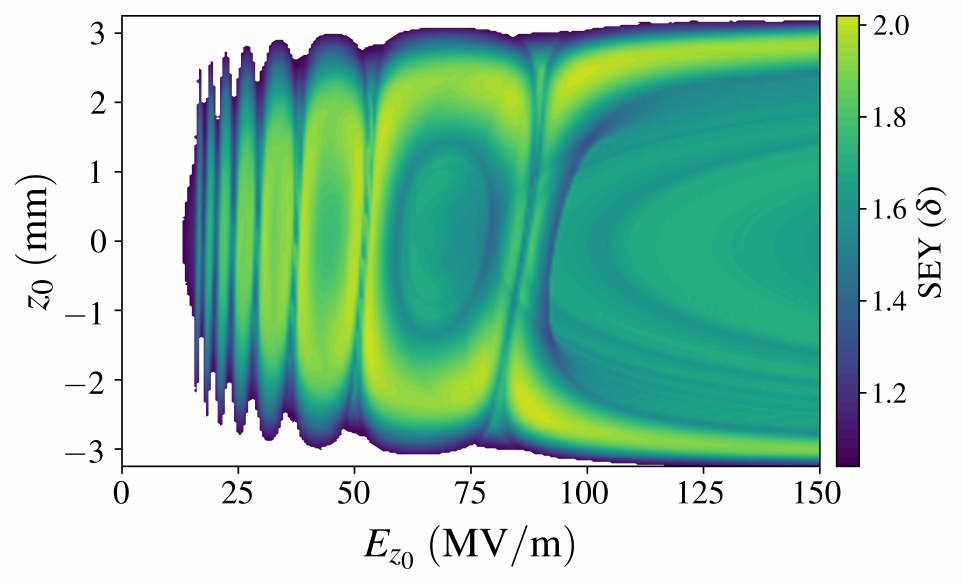}
\caption{\xl{Single-trip} SEY ($\delta$) map for the full cell of the $X$-band photogun cavities for different combinations of initial axial emission position \( z_0 \) and RF field amplitudes, characterized by the axial electric field at the cavity center \( E_{z0} \). The color scale indicates the magnitude of \( \delta \).}
    \label{fig:SEY_fullcell}
\end{figure}

To compute the SEY, we begin by extracting the impact energy \(W_i\) and incident angle \(\theta_i\) from the electron trajectory. The axial impact location is denoted by \(z_i\), and the RF phase at impact is given by \(\phi_i = \omega \tau + \phi_0\), where \(\tau\) is the time of flight. We assume an initial kinetic energy of 2~eV for secondary electrons, consistent with the emission model in Ref.~\cite{Vaughan1988}. We then perform a parameter sweep over the electron emission angle \(\alpha_0\) and initial RF phase \(\phi_0\), across a range of axial emission positions \(z_0\) and field amplitudes characterized by the axial electric field amplitude at the cavity center, \(E_{z0}\). All the field components (\(E_z\), \(E_r\), and \(B_\theta\)) are scaled proportionally with \(E_{z0}\), which serves as the reference parameter for the sweep. SEY values exceeding unity are retained and averaged over both $\alpha_0$ and $\phi_0$ to generate a representative \xl{single-trip} SEY contour map. The resulting map is shown in Fig.~\ref{fig:SEY_fullcell}, highlighting regions in \(E_{z0}\)–\(z_0\) space that are susceptible to enhanced secondary emission and the potential onset of electron multipacting.

\begin{figure}[t]
    \centering
    \includegraphics[width=\columnwidth]{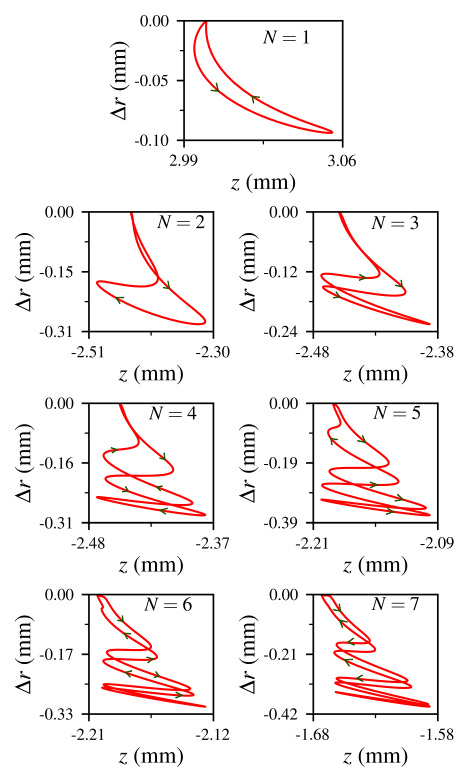}
\caption{Sample multipacting resonance modes (\( N = 1\text{--}7 \)) calculated for the RF field distribution in the full cell of the $X$-band photocathode cavities. The corresponding axial field amplitudes \( E_{z0} \) at the cavity center for each order are 106, 61, 39, 31, 26, 22, and 19~MV/m, respectively. \xlu{Arrows indicate the direction of electron motion along the trajectories.}}
    \label{fig:analyticalfullcell}
\end{figure}

\subsection{SEY under multipacting resonance}\label{sec:multi-resonance}
We now turn to the evaluation of SEY under multipacting resonance conditions. Multipacting resonance occurs when secondary electrons emitted from a surface are accelerated by an RF field such that their time of flight $\tau$ is synchronized with the RF period $T$~\cite{Shemelin2013}. The resonance order $N$ is given by $N = \tau / T$. When the SEY associated with a given resonance exceeds unity, the number of secondary electrons can increase with each RF cycle, potentially leading to an electron avalanche during a long RF pulse. \xl{Here we focus on converged resonant trajectories, i.e., trajectories that are identical across successive iterations. Identification of these resonant trajectories constitutes one original contribution of the analytical theory reported here, relative to Ref.~\cite{Haoran_2019}. These converged resonant trajectories are phase-locked and therefore more likely to contribute to dark current growth.}

To identify resonance conditions, we \xl{ fix one parameter, such as the emission angle $\alpha_0$, and calculate} the electron trajectory \xl{while sweeping the initial RF phase \( \phi_0 \) and axial emission position \( z_0 \) until convergence is reached}. Convergence is defined by the \xl{return} RF phase \( \phi \), axial position \( z \) and the SEY $\delta$ stabilizing within \( 1 \times 10^{-6} \) over ten consecutive iterations. \xl{An example of this convergence search process is presented in Appendix \ref{sec:conv}.}

\begin{figure}[t]
    \centering
    \includegraphics[width=\columnwidth]{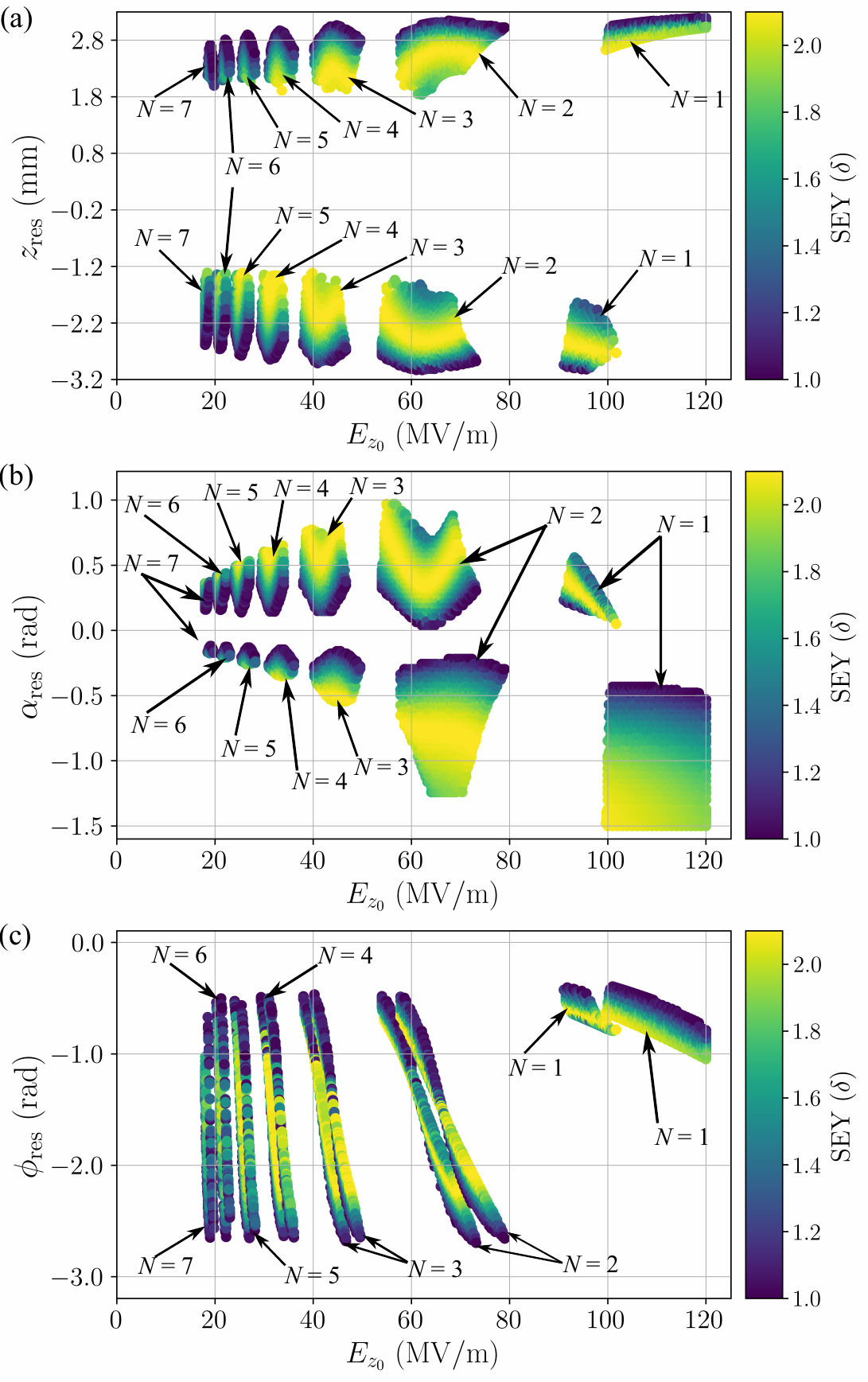}
    \caption{Multipacting susceptibility plots at resonance showing the SEY $\delta$ at various $E_{z0}$ with respect to (a) resonant axial emission position $z_{\text{res}}$, (b) resonant emission angle $\alpha_{\text{res}}$, and (c) resonant emission RF phase $\phi_{\text{res}}$. \xl{Each panel is a 2D projection of the same resonance islands, with the unplotted variables varying self-consistently to satisfy the resonance conditions.}
    }
    \label{fig:resonance_scatters}
\end{figure}

Figure~\ref{fig:analyticalfullcell} shows trajectories of multipacting resonance modes in the \( \Delta r\text{--}z \) plane for various resonance orders \( N \). SEY values are calculated for each mode across a range of initial conditions. The corresponding susceptibility map in Fig.~\ref{fig:resonance_scatters} plots SEY \( \delta \) as a function of \( E_{z0} \) and three key parameters: (a) resonant axial position \( z_{\text{res}} \), (b) emission angle \( \alpha_{\text{res}} \), and (c) emission phase \( \phi_{\text{res}} \). \xl{Figure~\ref{fig:resonance_scatters}(a)-(c) show different 2D cuts of the same resonance islands in the four-dimensional space of ($z_\text{res}$, $\alpha_\text{res}$, $\phi_\text{res}$, $E_{z0}$). In each panel, the other variables adjust self-consistently to satisfy the resonance conditions, rather than being held fixed.} These plots reveal distinct bands of sustained multipacting. As \( E_{z0} \) increases, dominant resonance orders shift from higher (\( N = 7 \)) to lower (\( N = 1 \)), a typical trend in RF multipacting behavior. \xlu{We note that, although these resonance bands shift slightly with the assumed initial energy of the secondary electrons (2~eV in the analytical model here), the range of field gradients that support multipacting resonances remains similar. This is also confirmed by the gradients at which multipacting current growth is predicted in the PIC simulations described in Sec.~\ref{sec:pic_mp}, where the emitted secondary electrons are modeled with the Vaughan energy spectrum.} Identifying these bands provides insight into multipacting onset and informs suppression strategies.

This analysis assumes steady-state RF fields. In short-pulse operation, transient fields limit the time available for multipacting to develop, thereby suppressing secondary electron growth. The next section presents time-domain simulations of dark current dynamics in the short-pulse regime.

\section{Dark Current Simulations}\label{sec:sims}
To investigate dark current behavior under short-pulse RF excitation, we perform numerical simulations using the \textsc{cst} Particle-In-Cell (PIC) and particle tracking solvers. This complements the steady-state analytical model by capturing transient effects in realistic cavity geometries. We begin by modeling field emission from high-gradient regions, such as the cavity iris and the photocathode surface. These field-emitted electrons can act as seed particles for multipacting-driven growth and subsequent dark current buildup, as discussed later.

\begin{figure}[t]
    \centering
    \includegraphics[width=\columnwidth]{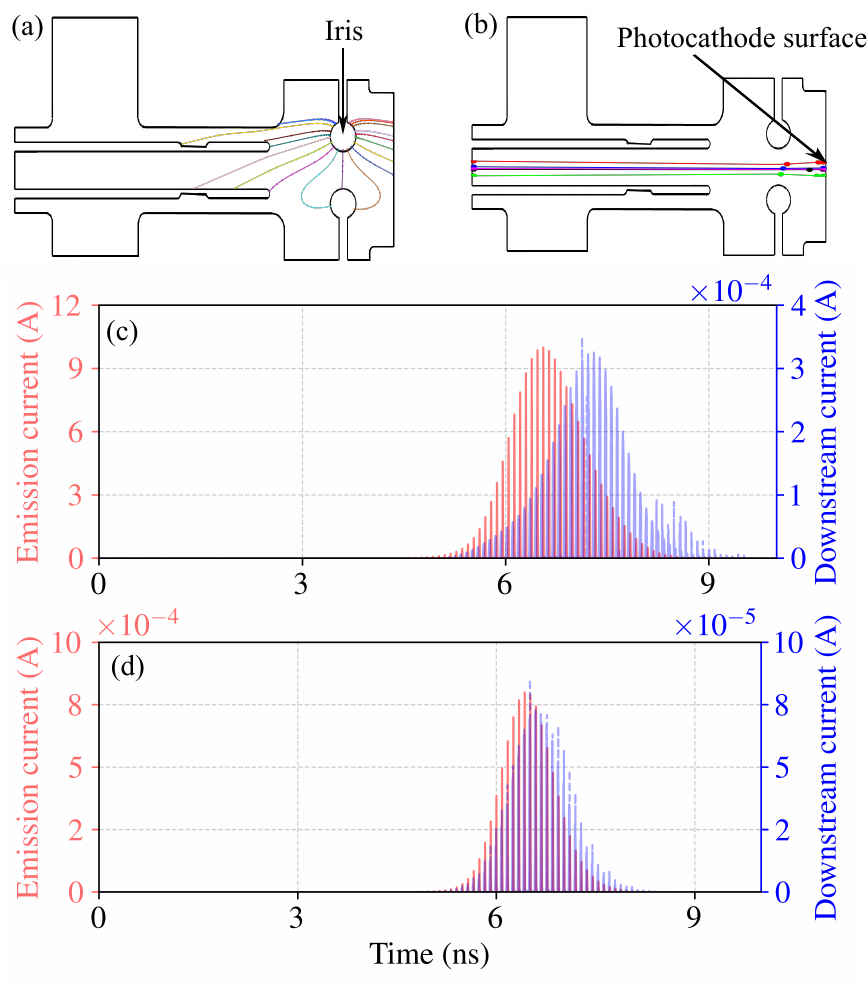}
   \caption{Simulated electron trajectories emitted from (a) the iris and (b) the photocathode surface, under a peak axial electric field amplitude of 325~MV/m at the center of the photocathode surface. Color-coded lines represent trajectories of electrons launched from different initial positions. \xl{Emitted current (red, left axis) and downstream current at the end of the beam pipe (blue, right axis; note the different scales) are shown in (c) for iris emission and in (d) for photocathode emission, assuming an estimated average field enhancement factor $\beta_\mathrm{avg} = 7$.}}
    \label{fig:tracks_iris_PC}
\end{figure}

\begin{figure}[t]
    \centering
    \includegraphics[width=0.95\columnwidth]{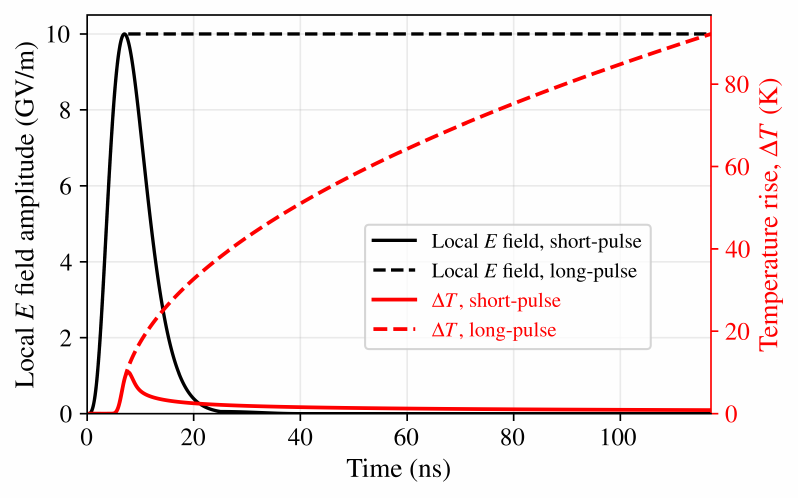}
    \caption{\xl{Temperature rise (\xlu{right} axis) due to field-emission-induced Joule heating, together with the local RF electric field amplitude (\xlu{left} axis). Results are shown for the 9 ns input RF pulse of Fig.~\ref{fig:Efield_combined}(c) and for a long pulse with the peak field extended by an additional 100 ns. Calculations assume a field emitter with a radius of 100~nm in a uniform RF field.}}
    \label{fig:heating}
\end{figure}

\subsection{Field Emission Simulations}\label{sec:FE}
Field emission is a critical process in RF cavities, where strong local electric fields extract electrons from metal surfaces. These electrons may disrupt the primary beam~\cite{Bane2005,ZHENG201512}, trigger RF breakdown~\cite{wang-1997-a}, or initiate dark current through multipacting and other processes. Extensive efforts have been made to model and characterize field emission across various RF structures~\cite{BIENVENU19921,Wu:2017sei,NOREM2005510,Huang2015_DarkCurrent}. 

Field emission simulations were performed using the \textsc{cst} PIC solver to interpret the low dark current observed during RF conditioning of the $X$-band photogun cavities.
% \xl{In these simulations, we adopted the Fowler-Nordheim equation field emission current density as:}
% \begin{align}
% \xl{j_F}
% &\xl{=}
% \xl{\frac{1.54\times10^{-6}\,10^{\,4.52\,\Phi^{-0.5}}}{\Phi}\;
% \beta^{2} E^{2}}
% \notag\\[-0.15em]
% &\xl{\quad\times
% \exp\!\left(
% -\frac{6.53\times10^{9}\,\Phi^{1.5}}{\beta\,E}
% \right),}
% \end{align}
% \xl{where $\Phi$ in eV is the work function of copper, as 4.65~eV; $\beta E$ is the local surface electric field with enhancement.}
As discussed in Sec.~\ref{sec:introduction}, no dark current signal was detected above the noise threshold~\cite{tan-2022}. This can be attributed to two factors: (1) the short RF pulse limits the total emitted charge, and (2) emitted electrons must traverse a long, narrow beam pipe (the inner conductor of the coaxial coupler) to reach diagnostics~\cite{Rijal:2024IPAC}. We illustrate the trajectories of field-emitted electrons in Fig.~\ref{fig:tracks_iris_PC}: in (a), most electrons emitted from the iris strike nearby surfaces; in (b), only electrons emitted near the center of the photocathode and within a narrow angular range are able to exit the beam pipe. \xl{To quantify dark current transmission, Figs.~\ref{fig:tracks_iris_PC}(c) and (d) plot the field-emitted current at the source and the downstream current measured at the end of the beam pipe for emission at the iris and at the photocathode, respectively. In these simulations the RF power is turned on at $t=0$. As noted earlier, no transmitted dark current was detected experimentally~\cite{tan-2022}, so we use this non-detection \xlu{(implying a per-pulse transmitted charge $< 1~\text{pC}$ from all dark-current mechanisms)} as an upper bound on $\beta$. \xlu{Field emission simulations indicate that an average $\beta_{\mathrm{avg}} \approx 8$ would yield a transmitted dark-current charge of $1~\text{pC}$ (field emission only); therefore, we adopt $\beta_{\mathrm{avg}} = 7$ as the representative case for the simulations in this section.} This $\beta_{\mathrm{avg}}$ choice when the entire surface is assumed to emit is equivalent, for the same peak field-emission current, to $\beta \approx 33.5$ assuming an active emitting area of a fraction of $10^{-9}$ of the surface.
}

\xl{Field emission can induce Joule heating, but the shorter emission window under short-pulse operation reduces energy deposition. We model a circular copper emitter with a radius of 100~nm and a uniform local RF electric field with a peak amplitude $\beta E = 10$~GV/m. The heating power density is calculated as $j_F^2/\sigma$\xlu{, where $j_F$ is the Fowler–Nordheim field-emission current density and} $\sigma$ is the electrical conductivity. We estimate the temperature rise using a semi-infinite copper half space~\cite{book-thermal}:}
\begin{equation}
\xl{\Delta T(0,t)=\frac{\sqrt{\alpha}}{\kappa \sqrt{\pi}}\int_{0}^{t}\frac{Q(\tau)}{\sqrt{t-\tau}}\,\mathrm{d}\tau,}
\label{eq:semiinf-convolution}
\end{equation}
\xl{where \(\kappa\), \(\rho\), and \(c_p\) are the thermal conductivity, mass density, and specific heat, respectively; $\alpha=\kappa/(\rho c_p)$ is the thermal diffusivity; and \(Q(\tau)\) is the heat flux \xlu{at time $\tau$}. Figure~\ref{fig:heating} compares the temperature rise in the short- and long-pulse cases. At the same peak field, the short pulse produces a distinctly lower $\Delta T$ than the long flat-top pulse. Therefore, short-pulse operation\xlu{, particularly with a short flat top,} mitigates field-emission-induced heating and increases the thermal margin for high-gradient operation.}

\begin{figure}[t]
    \centering
    \includegraphics[width=0.6\columnwidth, keepaspectratio]{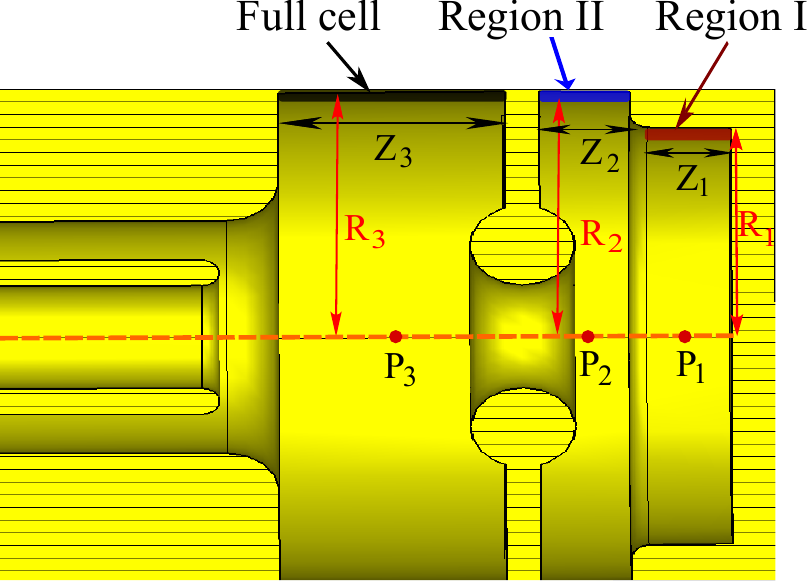}
    \caption{Multipacting simulation setup at the middle cutting plane of the $X$-band cavities, divided into the full cell, and Region I and Region II in the half cell. The reference points $P_1$, $P_2$ and $P_3$ denote the centers of each cavity, respectively. The relevant dimensions are: $R$\textsubscript{1} = 8.09~mm, $R$\textsubscript{2} = $R$\textsubscript{3} = 9.49~mm, $Z$\textsubscript{1} = 2.40~mm, $Z$\textsubscript{2} = 2.60~mm, and $Z$\textsubscript{3} = 6.50~mm.}
    % \xl{The highlighted area in orange defines the collision area used for the field emission induced Joule heating calculations in Sec.~\ref{sec:FE}.}}
    \label{fig:trajectory_setup}
\end{figure}

\xl{Field emission can also contribute to beam loading, but for experimentally relevant $\beta_{\mathrm{avg}}$ values, its direct effect is negligible, as verified with \textsc{cst} PIC simulations. It may still seed multipacting, which can then lead to loading.} Field emission transmission simulations in Fig.~\ref{fig:tracks_iris_PC} show that only about $10^{-5}$ of the total electrons emitted from the iris reach the end of the beam pipe. The electrons that strike cavity surfaces may initiate multipacting when resonance conditions are met.

\begin{figure}[t!]
    \centering
    \includegraphics[width=0.9\columnwidth]{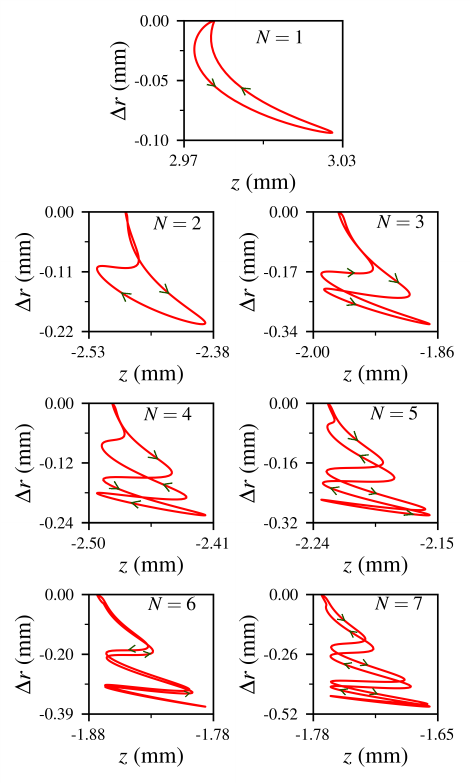}
    \caption{Sample multipacting resonance trajectories simulated in the full cell (defined in Fig.~\ref{fig:trajectory_setup}). The corresponding axial field amplitudes \( E_{z0} \) for resonance orders \( N = 1\text{--}7 \), evaluated at point \( P_3 \) (the center of the full cell), are 106, 61, 39, 31, 26, 22, and 19~MV/m, respectively. These trajectories show good agreement with the analytical results in Fig.~\ref{fig:analyticalfullcell}, as the RF field distribution in the full cell is well approximated by the analytical model.}
    \label{fig:trajfullcell}
\end{figure}

\begin{figure}[htpb!]
    \centering
    \includegraphics[width=0.9\columnwidth, keepaspectratio]{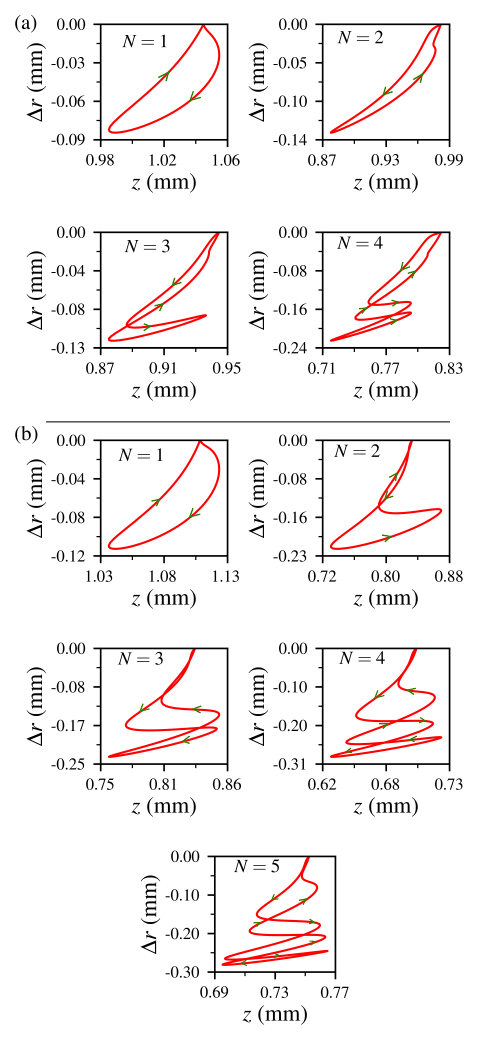}
\caption{Sample multipacting resonance trajectories in the half-cell for (a) Region~I and (b) Region~II. In Region~I, the axial fields \( E_{z0} \) evaluated at point \( P_1 \) for resonance orders \( N = 1\text{--}4 \) are 112, 61, 39, and 28~MV/m, respectively. In Region~II, the axial fields \( E_{z0} \) evaluated at point \( P_2 \) for resonance orders \( N = 1\text{--}5 \) are 95, 46, 30, 25, and 21~MV/m, respectively.
}
    \label{fig:traj_combined}
\end{figure}

\subsection{Multipacting simulations}\label{sec:pic_mp}
Multipacting behavior was simulated in \textsc{cst} using the realistic geometry of the $X$-band photocathode cavities driven by short RF pulses. Two types of simulations were performed, following the previous analytical studies: particle tracking simulations to identify multipacting resonance conditions by analyzing individual electron trajectories, and PIC simulations to evaluate dark current growth. 

In particle tracking simulations, electron trajectories were analyzed for emission from the sidewalls of the three regions as defined in Fig.~\ref{fig:trajectory_setup}, the full cell, and Region I and Region II in the half cell. Sample trajectories satisfying multipacting resonance conditions are presented in Figs.~\ref{fig:trajfullcell} and \ref{fig:traj_combined}. The resonant trajectories and the associated electric field amplitudes in the full cell (Fig.~\ref{fig:trajfullcell}) exhibit good agreement with the analytical results shown in Fig.~\ref{fig:analyticalfullcell} due to the similarity in the RF field pattern, providing further validation of the analytical model.

\begin{figure}[t]
    \centering
    \includegraphics[width=0.9\columnwidth]{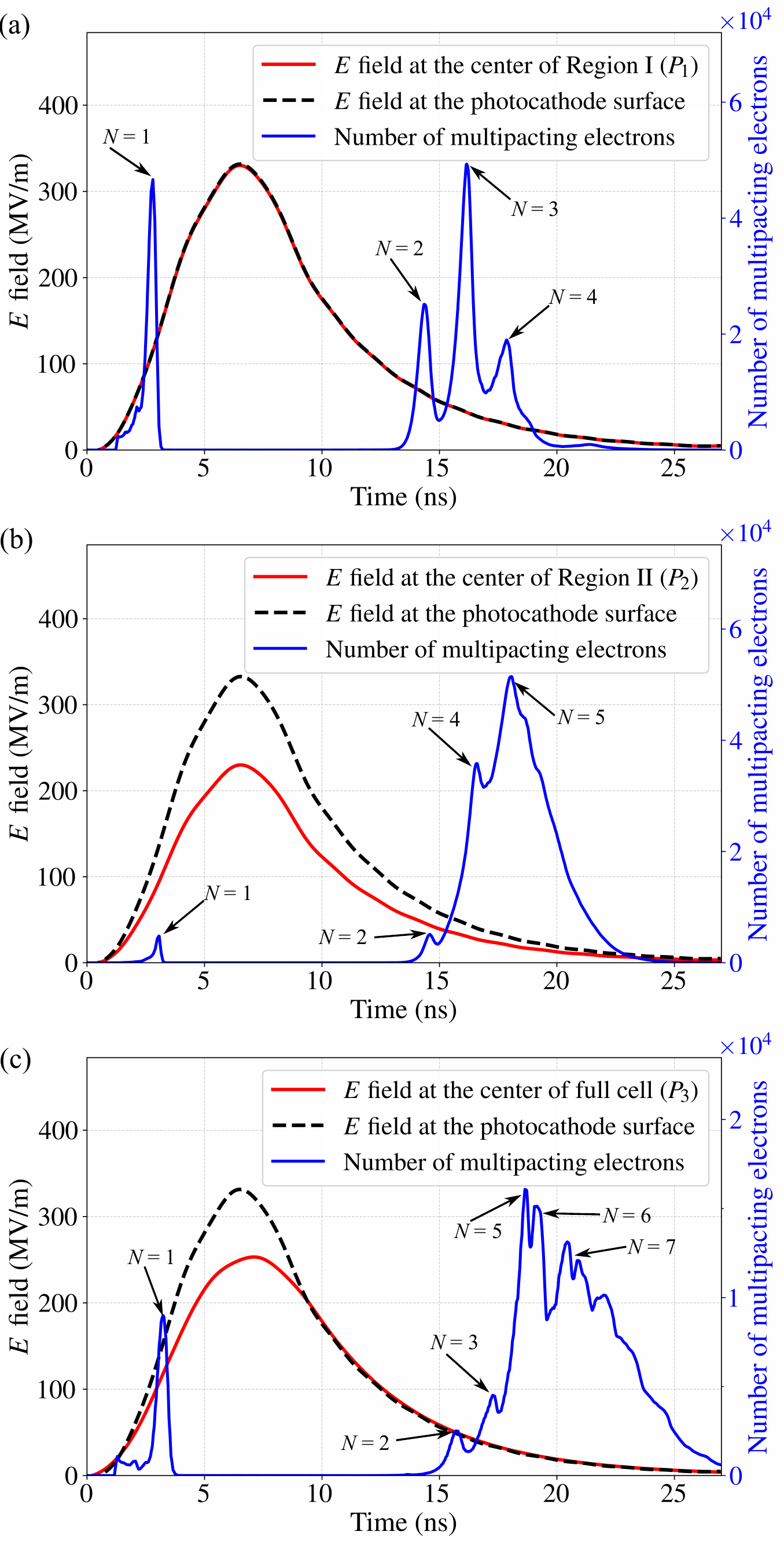}
    \caption{Temporal evolution of the electric field amplitudes and the number of multipacting electrons for \xl{simulations with the secondary electron emission surface defined in} (a) Region I, (b) Region II, and (c) the full cell. In each subplot, the on-axis electric field at point \( P_1 \), \( P_2 \), or \( P_3 \) (red solid line), the electric field across the photocathode surface (black dashed line), and the number of multipacting electrons (blue solid line) are shown. For Region~I, the dark current peaks corresponding to \( N = 1\text{--}4 \) appear at 113, 66, 43, and 30~MV/m; for Region~II, the peaks for \( N = 1, 2, 4, 5 \) are at 100, 47, 28, and 20~MV/m; and for the full cell, the peaks for \( N = 1, 2, 3, 5 \) are at 113, 51, 35, and 27~MV/m. Higher-order resonance modes (\( N \geq 6 \)) also contribute to later peaks in the full cell.}
    \label{fig:panel3MPCS}
\end{figure}

\begin{figure}[h]
    \centering
    \includegraphics[width=\linewidth]{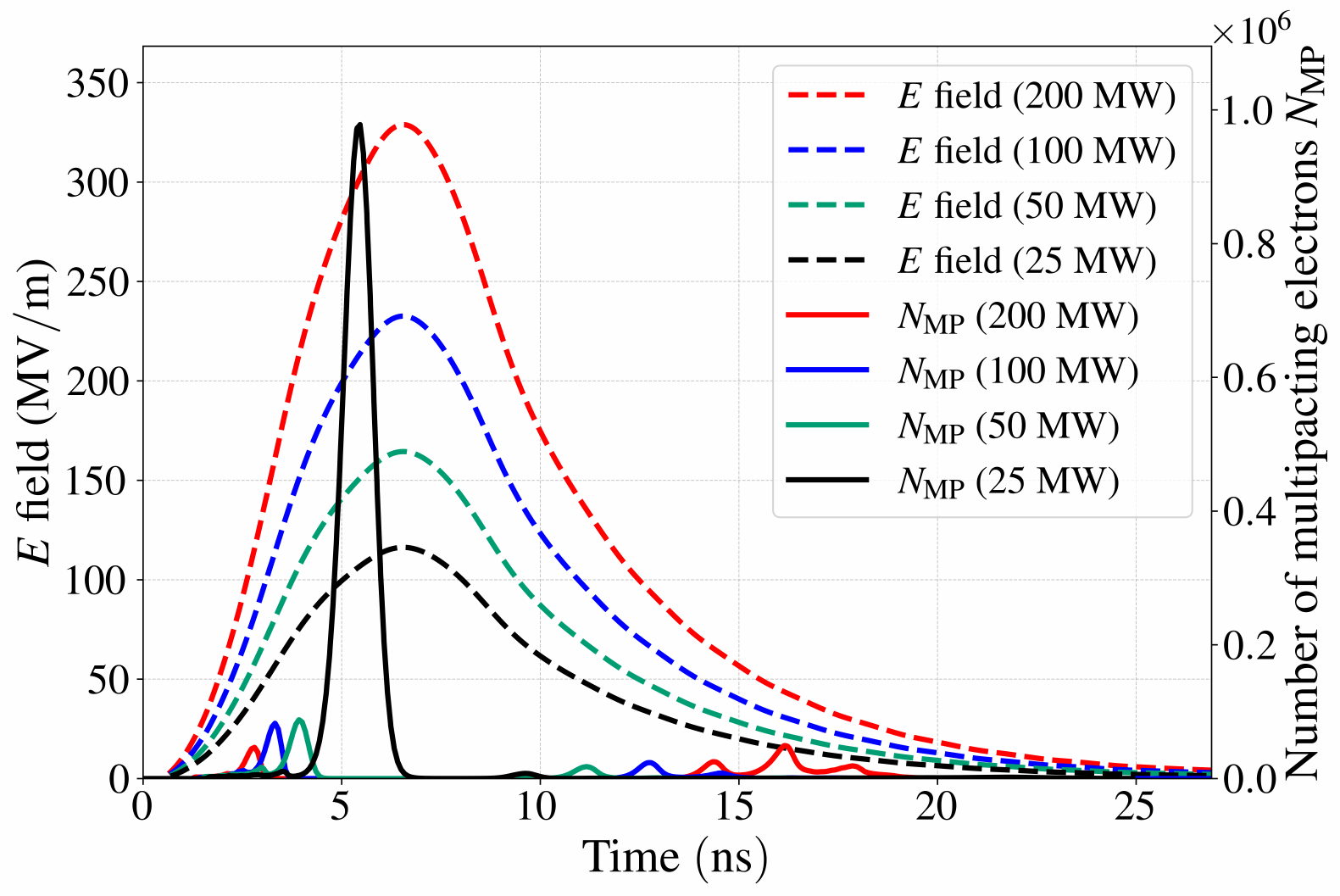}
    \caption{Number of multipacting electrons in Region I for RF pulses with a time profile of 3~ns rise, 3~ns flat top, and 3~ns fall at various peak power levels: 200 MW (red), 100 MW (blue), 50 MW (cyan), and 25 MW (black). Dashed curves represent the axial electric field amplitude at the center of Region I ($P_1$)
    while the corresponding solid curves in matching colors represent the number of multipacting electrons $N_{\text{MP}}$.}
    \label{fig:power}
\end{figure}

\begin{figure}[htpb!]
    \centering
    \includegraphics[width=\linewidth]{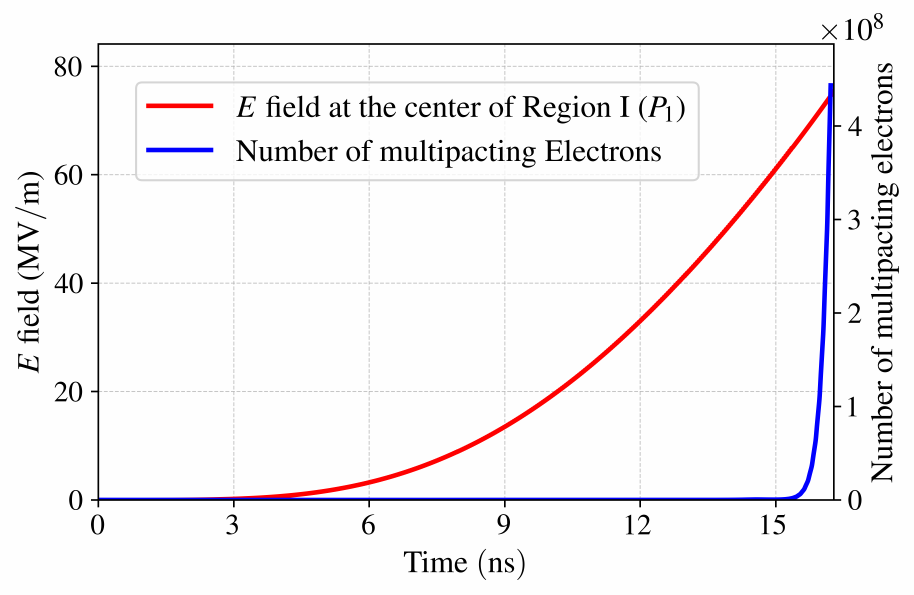}
    \caption{Temporal evolution of the electric field (red) and number of multipacting electrons (blue) using a longer RF pulse with a 30~ns rise from 0 to 200~MW. The simulation was terminated when the multipacting electron count exhausted available memory.} 
    \label{fig:longpulseMPC}
\end{figure}

In PIC simulations, we use again the short input RF pulse with a 3~ns rise, 3~ns flat top, and 3~ns fall and a peak power of 200~MW. Seed electrons were introduced \xl{at a rate of 400 electrons per RF cycle for 250 consecutive RF cycles,} using a procedure similar to those described in described \xl{in Refs.~\cite{Haoran_2019,zhang-2022}}. \xl{When these seed electrons collide with a surface without multipacting properties, they are removed from the simulation domain upon impact. When they collide with a surface designated for multipacting,} secondary emission is modeled using the Vaughan formalism~\cite{Vaughan1993}, with a maximum yield $\delta_{\text{max}} = 2.1$ at an impact energy $W_i = 150$~eV. \xlu{The emitted-energy spectrum is modeled with the \textsc{cst} copper parameterization~\cite{CST}.} Figure~\ref{fig:panel3MPCS} shows the temporal evolution of the electric field and the number of multipacting electrons in the three regions, \xl{where the secondary emission surface is on the sidewalls of Region I, Region II, or the full cell, respectively. The reported count is the total number of multipacting electrons in the simulation domain.} Peaks in the electron population correspond to distinct multipacting resonance orders, and the gradient at resonance are in strong agreement with the multipactor resonance trajectory analysis, confirming that electron growth is driven by multipacting resonances at the cavity sidewalls. \xl{Under the same input RF pulse, multipacting simulations at the iris and the photocathode surface show no sustained resonant growth of dark electrons.}

To further confirm that dark current growth arises from multipacting resonances, we varied the peak power of the input RF pulses while maintaining a fixed temporal profile. The resulting time evolution of the number of multipacting electrons is shown in Fig.~\ref{fig:power}. For a given cavity geometry, the field gradients at which multipacting resonance occurs are fixed. Adjusting the input power changes the time at which these gradients are reached, thereby shifting the onset of the resonance modes. The dark current peaks consistently occur at the same axial field amplitudes across different power levels, matching the resonance orders identified earlier. Each mode persists for a duration that depends on the field ramp rate. At lower gradients, the field ramp rate is reduced, so the peak dark current is higher due to the longer time window.

Short RF pulses can reduce multipacting dark current\xlu{, with short rise and fall times being particularly effective in limiting the time window for multipacting resonances.} Figure~\ref{fig:longpulseMPC} shows the simulation results with longer RF pulses as a comparison, where the gradual rise in field amplitude allows resonance conditions to persist longer, resulting in rapid growth of the multipacting electron population. In contrast, short RF pulses with short rise and fall times sweep through each resonance condition more quickly, reducing the duration over which the field remains within the amplitude window required for sustained multipacting and thus suppressing the buildup of secondary electrons. Therefore, the short-pulse regime reduces dark current by limiting the time available for resonance-driven electron multipacting.

\section{Impact of dark electrons}\label{sec:impact}
As dark electrons accumulate through field emission and multipacting, electron clouds can form. To observe this process, we perform \textsc{cst} PIC simulations with the cavity operated at the multipacting resonance conditions identified above. Seed electrons are briefly introduced near the cavity sidewall to initiate secondary emission. When the field satisfies resonance conditions, the electron population grows exponentially, forming a localized cloud near the sidewall.

\begin{figure}[t]
    \centering
    \includegraphics[width=\linewidth]{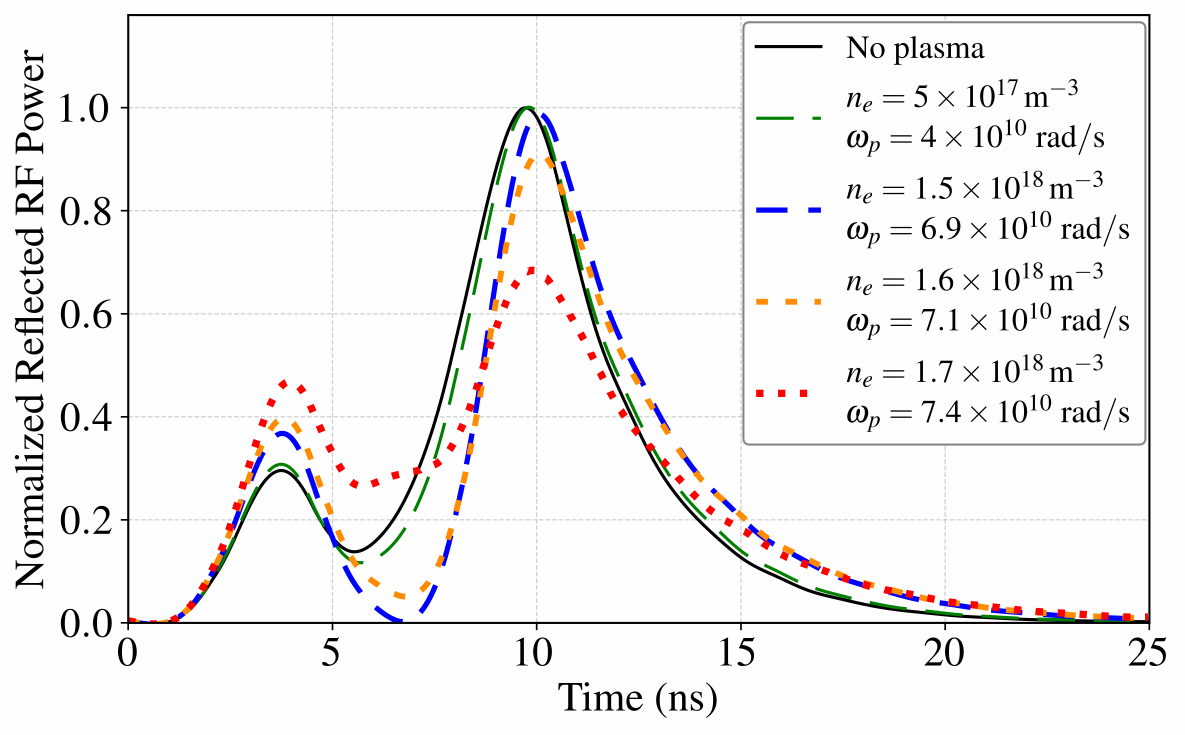} 
    \caption{Reflected RF power as a function of time for different plasma electron densities \( n_e \) \xlu{and the corresponding plasma frequencies $\omega_p$}, with the plasma modeled as a 0.2~mm thick layer on the cavity sidewall. All cases use the same short RF input pulse, and the traces are normalized to the no-plasma case.} 
    \label{fig:dispersion}
\end{figure}

A critical consequence of multipacting-induced electron clouds is dark-current-induced beam loading. To model this effect, we approximate the electron cloud as a cold plasma layer with a thickness of 0.2~mm, located along the cavity sidewall. \xl{The thickness is set to the average maximum wall-normal excursion of the multipacting trajectories (see Figs.~\ref{fig:trajfullcell} and \ref{fig:traj_combined}). Given the short pulse duration considered here, ion motion is negligible, with the ion velocity estimated to be on the order of $10^4$~m/s~\cite{johnson,fayawang_2009}.} The plasma response in this thin layer is described using the Drude dispersion, appropriate for an electron-only, collisionless medium. The effective permittivity of the layer is determined by an assumed plasma electron density \( n_e \). Figure~\ref{fig:dispersion} shows the normalized reflected RF power for different plasma densities, \xl{where the corresponding plasma frequency $\omega_p=\sqrt{n_e e^2/(\varepsilon_0 m)}$ is also displayed ($m$ is the electron mass).} As \( n_e \) increases, the reflected power decreases due to greater energy absorption by the plasma~\cite{Wuensch:2002vm}. This behavior is consistent with the experimentally observed drop in reflected power during conditioning of the $X$-band photocathode cavities (see Fig.~\ref{fig:conditioning}).

\section{Conclusion and discussion}\label{sec:conclusions}
Recent advances in short-pulse acceleration have opened new opportunities for achieving ultrahigh gradients, but they also require examination of the underlying RF breakdown mechanisms on previously underexplored timescales. Traditional breakdown studies, grounded in long-pulse operation and steady-state field assumptions, fall short in capturing the transient physics \xlu{characteristic of} nanosecond-long RF pulses.

In this work, we use analytical and numerical methods to show that RF pulses on the order of a few nanoseconds can effectively suppress dark current growth \xlu{by reducing contributions from mechanisms such as field emission, electron multipacting, and plasma formation. A short flat-top duration is essential for mitigating the impact of field emission by reducing the total emitted charge and associated effects such as field-emission-induced Joule heating.} \xlu{For multipacting, our analytical and numerical models} identify resonance bands, while time-domain simulations reveal how rapid field ramping shortens the windows for sustained development of multipacting resonances. \xlu{Short ramp-up and ramp-down times are therefore crucial for suppressing multipacting by limiting the time available for resonance modes to develop. The reduced electron population also lowers the plasma density, mitigating its impact on the RF pulse.} These results provide strong evidence that short-pulse operation alters the onset dynamics of dark current and RF breakdown. Our simulation results are also consistent with experimental observations of low dark current and beam loading in the $X$-band photogun cavities at AWA.

To advance the understanding of RF breakdown in the short-pulse regime, future work should focus on modeling the dynamics of plasma formation and its interaction with time-varying RF fields, ideally using first-principles simulation tools. Conventional breakdown models often treat plasma evolution as a multi-stage process, with nanosecond-scale electron dynamics coupled to slower changes in surface morphology. However, in the short-pulse regime, where the electromagnetic fields vary rapidly in time, these transient interactions may dominate the breakdown initiation process. \xl{In particular, a self-consistent model that includes both field emission and multipacting, with high-performance computing resources, would enable direct comparison with the experimental setup and is a valuable direction.} Investigating these fast-timescale effects will be essential for refining our understanding of the coupled physics in RF breakdown, and short-pulse experiments provide a promising platform for isolating and studying these mechanisms.

\begin{acknowledgments}
The authors thank Drs. Chengkun Huang, Omkar Ramachandran and Haoran Xu for valuable discussions.
This research was supported by the U.S. Department of Energy,
Office of Science, Office of High Energy Physics under Award DE-SC0021928 and DE-SC0022010.
\end{acknowledgments}

\appendix
\section{Simplified theory of multipactor in crossed RF fields with closed-form solutions}~\label{sec:simplified}
We present a simplified theoretical framework to describe the multipacting process near the cavity sidewall. This analysis focuses on the essential dynamics governing electron motion under RF electric and magnetic fields. The key distinction between this simplified theory and the analytical model in Sec.~\ref{sec:analytical} is the neglect of the radial variation of \(E_r\) and \(B_\theta\), which allows for closed-form solutions of the multipacting resonance conditions. The schematic setup for this analysis is the same as that shown in Fig.~\ref{fig:combined_crossfield}.

We assume that \(E_r\) and \(B_\theta\) are independent of \(r\) near the sidewall, so \(E_r\) is expressed as:
\begin{equation}
E_{r}(z) = E_{r0} \cos\left(\frac{\pi z}{L}\right) \cos(\omega t + \phi_0),
\label{eq:radialEfield}
\end{equation}
where \(E_{r0}\) is the peak amplitude of \(E_r\), \(\omega\) is the angular frequency of the RF field, and \(\phi_0\) is the RF phase. The azimuthal magnetic field \(B_\theta\) is approximated as:
\begin{equation}
B_\theta (z)= \left[ B_0 + B_1 \sin\left( \frac{\pi z}{L} + \zeta \right) \right] \sin(\omega t + \phi_0),
\label{eq:magnetic_field_wall}
\end{equation}
where the coefficients \(E_{r0}\), \(B_0\), \(B_1\), and \(\zeta\) are determined by fitting to \textsc{cst} simulation results of the field distribution near the sidewall.
The axial electric field \(E_z(r)\) retains its radial dependence and is written as:
\begin{equation}
E_z(r) = E_{z0} P \Delta r \cos(\omega t + \phi_0)
\label{eq:axialEfield}
\end{equation}
Here, \( E_{z0} \) denotes the axial field amplitude at the cavity center, and \( \Delta r = r - a \) is the distance from the sidewall. The local radial derivative of the axial field is defined as \( P = \frac{dE_z/dr}{E_{z0}} \), which, under Maxwell's equations, can be approximated as \( P = \frac{i \omega B_\theta}{E_{z0}} \). Under this approximation, the radial electric field component \(E_r\) can be neglected in the equation for the axial velocity \(v_z\) (see also Ref.~\cite{Shemelin2013}).

\begin{figure}[t!]
    \centering
    \includegraphics[width=\columnwidth]{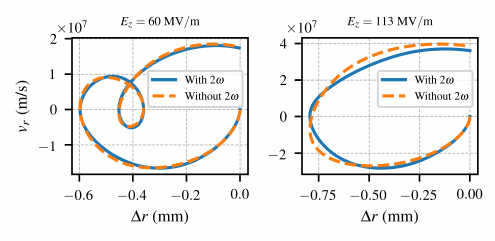}
    \caption{Phase-space trajectories for $N=2$, $E_{z0} = 60\ \mathrm{MV/m}$ (left) and $N=1$, $E_{z0} = 113\ \mathrm{MV/m}$ (right) cases with and without the $2\omega$ term in Eq.~(\ref{eq:radial_acc_equation}). The approximation shows good agreement for lower fields when $N\geq2$, but diverges for $N=1$ at high fields.}
    \label{fig:phase_space_comparison}
\end{figure}

Using the field distribution given in Eqs.~\eqref{eq:radialEfield}--\eqref{eq:axialEfield}, we substitute them into the equations of motion:
\begin{equation}\label{eq:axial_simplified}
    \frac{dv_z}{dt} = -\frac{e}{m} \frac{d}{dt} \left[ B_{\theta} \, \Delta r \sin(\omega t + \phi_0) \right].
\end{equation}
Integrating Eq.~\eqref{eq:axial_simplified} using the initial condition \(v_z(t = 0) = v_0 \sin(\alpha_0)\), we obtain \(v_z(t)\), which is then substituted into the equation for \(v_r\):
\begin{multline}
    \frac{dv_r}{dt} = -\frac{e}{m} E_r \cos(\omega t + \phi_0) 
    + \frac{e}{m} v_0 B_\theta \sin \alpha_0  \sin(\omega t + \phi_0) \\
    - \left( \frac{e}{m} \right)^2 B_\theta^2 \, \Delta r \frac{1 - \cos(2\omega t + 2\phi_0)}{2}.
    \label{eq:radial_acc_equation}
\end{multline}
The terms oscillating at \(2\omega\) are neglected, as their contribution is negligible in the low-field limit. Their influence becomes significant at higher fields, as demonstrated by the phase-space trajectories in Fig.~\ref{fig:phase_space_comparison}. In this example, the approximation is more accurate for multipacting orders $N\geq2$.

By omitting the \(2\omega\) terms, Eq.\eqref{eq:radial_acc_equation} simplifies to the following second-order differential equation:
\begin{multline}
    \frac{d^2 \Delta r}{dt^2} + \frac{1}{2} \left( \frac{e}{m} \right)^2 B_\theta^2 \Delta r = 
    -\frac{e}{m} E_r \cos(\omega t + \phi_0) \\
    + \frac{e}{m} v_0 B_\theta \sin \alpha_0 \sin(\omega t + \phi_0).
\end{multline}
Following the approach in Ref.~\cite{Miller1958}, this equation takes the form of a driven harmonic oscillator:
\begin{equation}
        \frac{d^2 \Delta r}{dt^2} + \Omega^2 \Delta r = f(\omega t),
\end{equation}
where \( f(\omega t) \) represents the RF driving term, and the characteristic frequency \(\Omega\) is defined as:
\begin{equation}
    \Omega = \frac{1}{\sqrt{2} m} e B_\theta.
\end{equation}
This solution holds under the condition \( \Omega \ll \omega \), but begins to deviate from full numerical results when \( \Omega \gtrsim 0.25\,\omega \). For the system studied here, the condition of \( \Omega = \omega \) occurs at \( E_{z0} = 260.6~\mathrm{MV/m} \), indicating the upper limit of validity for the approximation. \xl{Here $E_{z0}$ is computed using parameters obtained by fitting the simulated RF field distribution to Eq. (\ref{eq:magnetic_field_wall}).}

The multipacting resonance condition is then given by:
\begin{equation}
    \Omega N T = \pi,
    \label{eq:resonance_condition}
\end{equation}
which can be rearranged as:
\begin{equation}
    \quad \frac{e B_\theta}{m} = \frac{\omega}{N \sqrt{2}},
\end{equation}
where \( T = (2\pi)/\omega \) is the RF period and \( N \) is again the multipacting order. 

While this simplified approach neglects the radial variation of the transverse fields \( E_r \) and \( B_\theta \), and thus cannot predict the SEY with the same accuracy as the full analytical model, it offers a closed-form expression that effectively approximates multipacting resonance conditions, especially at low fields.
 
 \begin{table}[t]
\centering
\begin{ruledtabular}
\begin{tabular}{c|cc|cc}
\multicolumn{1}{c|}{\textbf{Multipacting}} & 
\multicolumn{2}{c|}{\textbf{Full Analytical}} & 
\multicolumn{2}{c}{\textbf{Simplified Analytical}} \\
\multicolumn{1}{c|}{\textbf{Resonance}} & 
\makecell{\textbf{$E_{z0}$ min}} & 
\makecell{\textbf{$E_{z0}$ max}} & 
\makecell{\textbf{$E_{z0}$ min}} & 
\makecell{\textbf{$E_{z0}$ max}} \\
\makecell{\textbf{Order ($N$)}} & 
\makecell{(MV/m)} & 
\makecell{(MV/m)} & 
\makecell{(MV/m)} & 
\makecell{(MV/m)} \\
\hline
1 & 91.1 & 120.0 & 116.0 & 149.0 \\
2 & 54.0 & 79.0  & 58.0  & 74.0  \\
3 & 38.0 & 49.5  & 39.0  & 49.0  \\
4 & 29.4 & 36.1  & 29.0  & 37.0  \\
5 & 24.0 & 28.2  & 24.0  & 29.0  \\
6 & 20.3 & 23.0  & 20.0  & 24.0  \\
7 & 18.0 & 19.8  & 17.0  & 21.0  \\
\end{tabular}
\end{ruledtabular}
\caption{Comparison of $E_{z0}$ ranges for various multipacting resonance orders $N$, calculated using the simplified and full analytical models. The simplified model captures the scaling trend well, particularly for higher-order resonances (\( N \geq 2 \)).}
\label{tab:ez0_singlecol}
\end{table}

Table~\ref{tab:ez0_singlecol} summarizes the predicted ranges of \( E_{z0} \) corresponding to various multipacting orders, as computed using both the simplified and full analytical models. These ranges are obtained by sweeping over the resonant parameters \( z_{\text{res}} \), \( \alpha_{\text{res}} \), and \( \phi_{\text{res}} \). Notably, good agreement is observed for higher-order multipacting modes (\( N \geq 2 \)), indicating that the simplified theory effectively captures the key scaling behavior.

\begin{figure}
  \centering
  \includegraphics[width=0.95\linewidth]{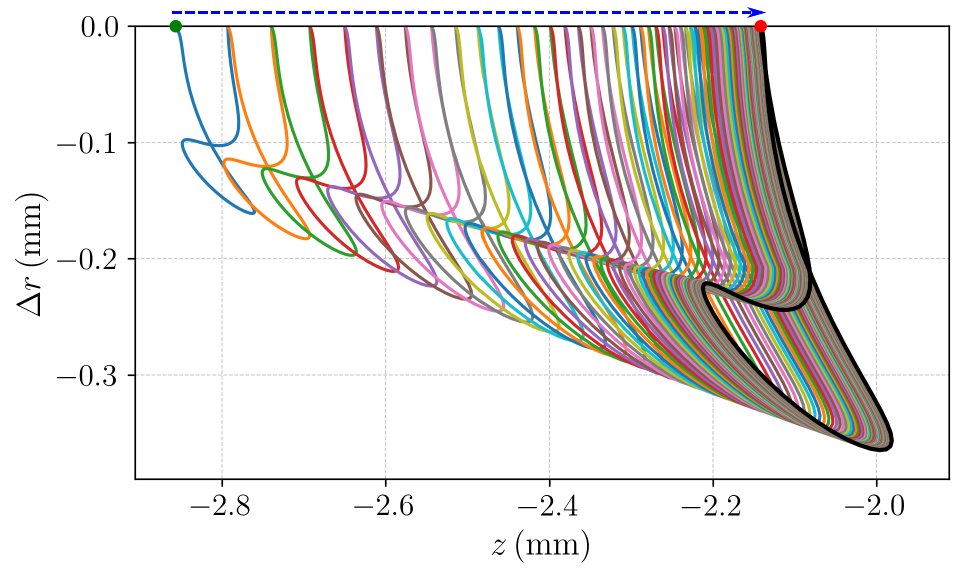}
  \caption{\xl{An example of convergence search for multipacting resonant trajectories with a fixed emission angle $\alpha =42.2^\circ$, showing $\Delta r$ versus $z$. The green dot marks the initial release point; the red dot marks the resonant point where the trajectory stabilizes after multiple iterations.}}
  \label{fig:iterations}
\end{figure} 

\begin{figure}
  \centering
  \includegraphics[width=0.8\linewidth]{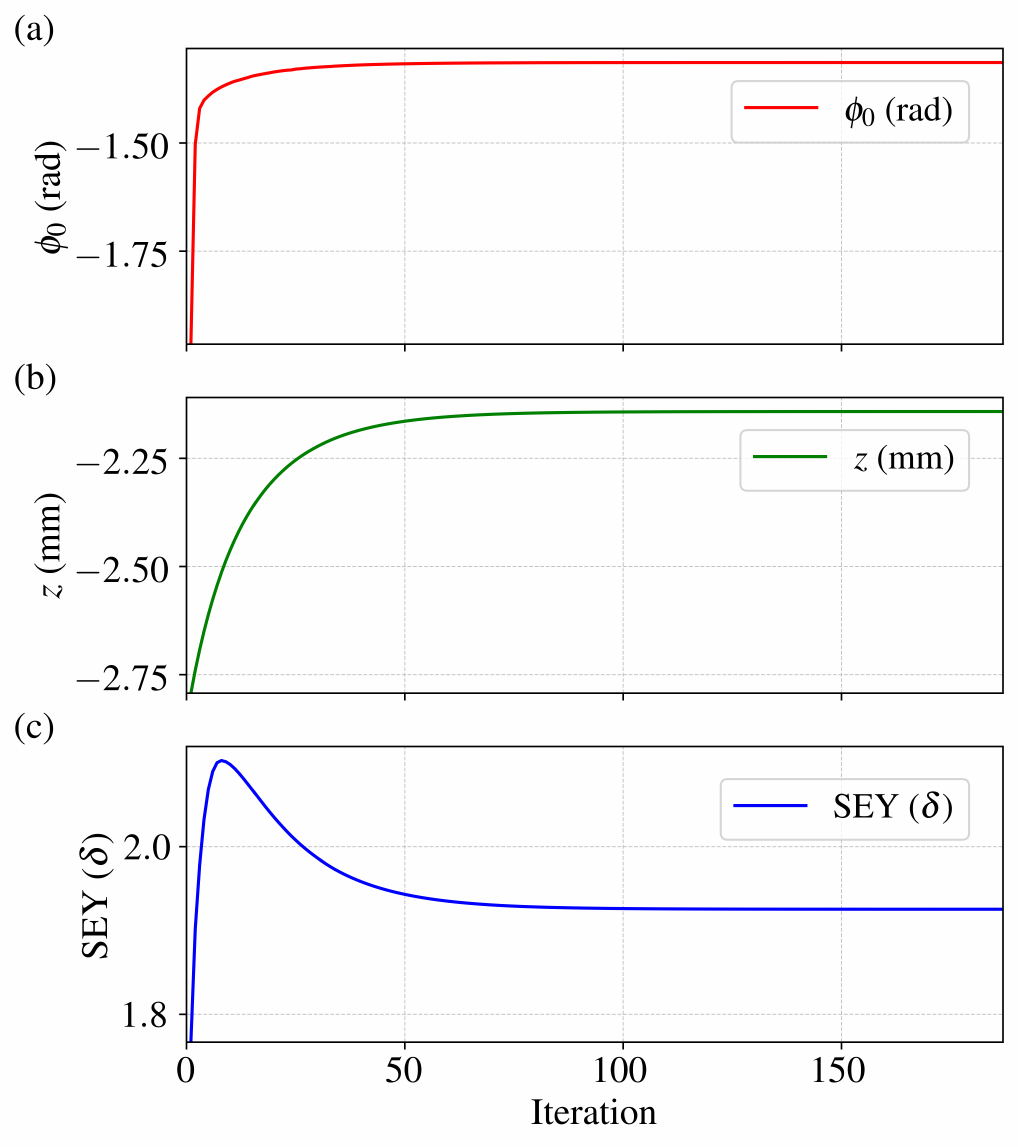}
  \caption{\xl{Convergence search over multiple iterations for the example shown in Fig.~\ref{fig:iterations}, demonstrating convergence to a resonant, phase-locked trajectory: (a) phase $\phi$, (b) axial position $z$, and (c) SEY $\delta$.}}
  % \vspace{-5mm}
  \label{fig:A3}
\end{figure}

\section{\xl{Convergence search of multipacting resonant trajectories}}\label{sec:conv}

\xl{As mentioned in Sec.~\ref{sec:multi-resonance}, the analytical theory in this work identifies converged and phase-locked multipacting resonant trajectories, in contrast to previous studies~\cite{Haoran_2019}. In this appendix, we present an example of the convergence search process, shown in Fig.~\ref{fig:iterations}. In this example, we fix the emission angle $\alpha$ while the other variables are allowed to vary. Each iteration integrates one trip from emission to impact, while the \xlu{RF phase $\phi$ and axial position $z$ at} emission are set to the previous impact values. The earlier paths are not resonant, while the trajectory ultimately converges to a phase-locked stable orbit. \xlu{These successive iterations do not represent the physical time evolution of secondary electron motion over multiple impacts; rather, they constitute a numerical procedure to locate phase-locked resonance conditions.} Convergence is declared when, for ten consecutive iterations, the changes in $\phi$, $z$, and SEY $\delta$ between iteration $k$ and $k+1$ each fall below $10^{-6}$, while $\delta$ remains greater than unity:}
\begin{multline*}
\xl{|\phi_{k+1}-\phi_k|<10^{-6},\quad |z_{k+1}-z_k|<10^{-6},}\\
\xl{|\delta_{k+1}-\delta_k|<10^{-6},\quad \delta_k>1.}
\end{multline*}
\xl{Figure~\ref{fig:A3} illustrates the convergence of these quantities, confirming the establishment of the resonance condition.}

\end{document}